\documentstyle[sprocl]{article}

\def\be{\begin{equation}}
\def\ee{\end{equation}}
\def\ba{\begin{array}}
\def\ea{\end{array}}

\def\L{\Lambda}
\def\l{\lambda}

\def\Nb{{I\!\! N}}

\def\Cb{\ \hbox{\vrule width 0.6pt height 6pt depth 0pt
              \hskip -3.2 pt} C}

\def\bru{\vert 1\rangle}
\def\brd{\vert 0\rangle}

\def\a{\alpha}
\def\b{\beta}
\def\d{\delta}

\def\g{\gamma}

\def\l{\lambda}
\def\o{\omega}
\def\p{\phi}

\def\r{{\cal \rho}}
\def\s{\sigma}
\def\x{\chi}

\def\L{\Lambda}
\def\P{\Psi}
\def\Q{\Phi}
\def\F{{\cal F}}

\def\uqA22{{U_q(A^{(2)}_2)}}

\def\beq{\begin{equation}}
\def\eeq{\end{equation}}
\def\bea{\begin{eqnarray}}
\def\eea{\end{eqnarray}}
\def\ba{\begin{array}}
\def\ea{\end{array}}
\def\no{\nonumber}
\def\lt{\left}
\def\rt{\right}
\newcommand{\bq}{\begin{quote}}
\newcommand{\eq}{\end{quote}}

\newtheorem{Theorem}{Theorem}

\newcommand\ket[1]{\left| #1\right\rangle}

\begin{document}

\title{Quantum Teleportation:\\ from Pure to Mixed States and Standard to Optimal}

\author{S. Albeverio}

\address{
Institut f\"ur Angewandte Mathematik,
Universit\"at Bonn, D-53115\\
SFB 611; IZKS; BiBos; CERFIM(Locarno)\\
E-mail: albeverio@uni-bonn.de}

\author{S.M. Fei}

\address{Institut f\"ur Angewandte Mathematik,
Universit\"at Bonn, D-53115\\
Department of Mathematics, Capital Normal
University, Beijing 100037\\
E-mail: fei@uni-bonn.de}

\author{W.L. Yang}

\address{
Institute of Modern Physics, Northwest University,
Xian 710069\\
Yukawa Institute for Theoretical Physics, Kyoto
University, Kyoto 606-8502\\
E-mail: wlyang@yukawa.kyoto-u.ac.jp}

\maketitle

\abstracts{Teleportation for pure states, mixed states with
standard and optimal protocols are introduced and investigated
systematically. An explicit equation governing the teleportation
of finite dimensional quantum pure states by a generally
given non-local entangled state is
presented. For the teleportation of a mixed state with an
arbitrary mixed state resource, an explicit expression is obtained
for the quantum channel associated with the standard teleportation
protocol. The corresponding transmission fidelity
is calculated. It is shown that the standard
teleportation protocol is not optimal. The optimal quantum
teleportation is further studied, its fidelity
is given and shown to be related to the fully
entangled fraction of the quantum resource, rather than the single
fraction as in the standard teleportation protocol.}

\section{Introduction}

One of the most profound results of quantum information theory is
the discovery of a quantum teleportation protocol \cite{Bennett93}. By
means of a classical communication channel and a quantum resource
realized by a nonlocal entangled state such as an EPR-pair of
particles, the teleportation process allows to transmit an unknown
quantum state from a sender traditionally named ``Alice" to a
receiver ``Bob" which are spatially separated.
Quantum teleportation has been introduced in [1]
and discussed by a number of authors for both spin-$\frac{1}{2}$
states and arbitrary quantum states, see e.g. [2-11].

For teleportation of $N$-dimensional quantum states, the
teleportation problem has been discussed in [7]
in the case where
the dimensions of the Hilbert spaces associated with the sender,
receiver and the auxiliary space are all equal to $N=2^{m}$,
for a given $m\in\Nb$. The relations among
quantum teleportation, dense coding,
orthonormal bases of maximally entangled vectors
and unitary operators with respect to the
Hilbert-Schmidt scalar product, and depolarizing operations
are investigated in [8].

These teleportation processes can be viewed as quantum channels.
The nature of a quantum channel is determined by the particular
protocol and the state used as a teleportation resource
\cite{Bennett93,Hor01,Ari00}. The standard teleportation protocol
$T_0$ proposed in [1] uses {\it Bell} measurements and
{\it Pauli} rotations. When the maximally entangled pure state
$|\Q\rangle=\frac{1}{\sqrt{n}}\sum_{i=0}^{n-1}|ii\rangle$ is used as the
quantum resource, it provides an ideal noiseless quantum channel
$\L^{(|\Q\rangle\langle\Q|)}_{T_0}(\r)=\r$. However in realistic situations,
instead of pure maximally entangled states, Alice and Bob
usually share a mixed entangled state due to the decoherence.
Teleportation using a mixed state as an entangled resource is, in
general, equivalent to having a noisy quantum channel. Recently,
an explicit expression for the output state of the quantum channel
associated with the standard teleportation protocol $T_0$ with an
arbitrary  mixed state resource has been obtained
\cite{Bow01,Alb02}.

In this article we give a systematic description of quantum
teleportations for general quantum states \cite{Alb02,Alb03,Alb04}. We first
discuss the general properties of teleportation for finite
dimensional quantum pure states without the assumption on equality
for the dimensions of the Hilbert spaces involved. We give a
teleportation protocol for generally given entangled states and a
constraint equation that governs the teleportation. The solutions
of the constraint equation give the unitary transformations of
teleportation protocols. Detailed examples and the roles played by
the dimensions of the Hilbert spaces are discussed.

We then consider the teleportation of mixed states. Using a
different approach as compared with [14] we derive an explicit
expression of the protocol for the quantum channel associated with
the standard teleportation protocol $T_0$ and an arbitrary mixed
state resource. We further calculate the transmission fidelity and
show that the transmission fidelity of the standard teleportation
protocol with an arbitrary mixed state $\x$ as a resource depends on
the singlet fraction of the resource $\x$.

To investigate the optimal teleportation, we consider the
following problem. Alice and Bob previously only share a pair of
particles in  an arbitrary mixed entangled state $\x$. In order to
teleport an unknown state $\r$ to Bob, Alice first performs a
joint Bell measurement on her particles (particle 1 and particle
2) and tell her result to Bob by the classical communication
channel. Then Bob, instead of the {\it Pauli} rotation like in the
standard teleportation protocol\cite{Bennett93}, tries his best to
choose a particular unitary transformation which depends on the
quantum resource $\x$, so as to get the maximal transmission
fidelity. We derive an explicit expression for the quantum channel
associated with the optimal teleportation with an arbitrary mixed
state resource. The transmission fidelity of the corresponding
quantum channel is given in terms of the fully entangled fraction
of the quantum resource.

\section{A most simple example -- teleportation of a qubit state}

We address the quantum teleportation problem from a most simple
example: the teleportation of a quantum bit (qubit) state. The
state of a qubit is mathematically a vector in two dimensional
complex Hilbert space ${\cal H}$. Taking two bases of ${\cal H}$ to be
$\bru=\left(\ba{l} 0\\1\ea\right)$  and $ \brd=\left(\ba{l}
1\\0\ea\right)$, a general state of a qubit  can be written as
\be\label{a1}
\vert\alpha\rangle=a\bru+b\brd=\left(\ba{l} a\\
b\ea\right),~~~a,~b\in \Cb
\ee
with normalization $\vert a\vert^2+\vert b\vert^2=1$.
A quantum measurement on $\phi$ would projects the state to
$\bru$ (resp. $\brd$) with probability $\vert a\vert^2$
(resp. $\vert b\vert^2$).

The states of multiple qubits are then the vectors on the space
${\cal H}\otimes {\cal H}\otimes...\otimes {\cal H}$.
A vector on ${\cal H}\otimes {\cal H}\otimes...\otimes {\cal H}$
that can not be written as
$(a_1\bru+b_1\brd)\otimes(a_2\bru+b_2\brd)
\otimes...\otimes(a_n\bru+b_n\brd)$ for some $a_i,~b_i\in\Cb$ is called
entangled. For example, the EPR (Einstein, Podolsky and Rosen) pair
\be\label{a2}
\beta=\frac{1}{\sqrt{2}}(\vert 00\rangle+\vert 11\rangle)
\ee
is a maximally entangled state.

The transformations on the vector states are called quantum gates.
As a quantum system evolves unitarily, these quantum gates are just
unitary transformations given by matrices $M$ such that $MM^\dag = I$,
where $\dag$ stands for conjugation and transpose, $I$ is the identity matrix.

The followings are typical quantum gates on a single qubit:
$$
I=\left(\begin{array}{cc} 1 ~&~  0\\   0 ~&~
1\end{array}\right),~~~
X=\left(\begin{array}{cc} 0 ~&~  1\\   1 ~&~
0\end{array}\right),~~~
Y=\left(\begin{array}{cc} 0 &  -1\\   1 &
0\end{array}\right),~~~
Z=\left(\begin{array}{cc} 1 &  0\\   0 &
-1\end{array}\right),
$$
where $X$ (resp. $Z$) functions as negation
(resp. phase shift), $Y = ZX$.
A useful gate on two qubits called
Controlled-{\sc not} gate is given by
$$
C_{not}=
\left(\begin{array}{cccc} 1 & ~0~ &~ 0~ & 0\\
 0 &~  1~ & ~0~ & 0\\  0 &~  0~ &~  0~ &  1\\
0 &  ~0 &~  1~ &  0\end{array}\right).
$$
It is easily checked that $C_{not}$ maps
$\ket{00} \to  \ket{00},~  \ket{01} \to  \ket{01},~
\ket{10}\to \ket{11}$, and $\ket{11} \to \ket{10}$.
Another useful gate is called Hadamard Transformation $H$:
$\displaystyle
H=\frac{1}{\sqrt{2}}\left(\ba{cc} 1& 1\\[2mm]  1&
-1\ea\right).
$
It maps
$
\ket{0}  \to
\displaystyle {1\over \sqrt 2}(\ket 0 + \ket
1)$ and
$\ket{1}  \to \displaystyle  {1\over \sqrt
2}(\ket 0 - \ket 1)$.

Now Alice has a general qubit state (\ref{a1}) which is unknown to her
(she does not know the parameters $a$ and $b$).
But she would like Bob to have a qubit state of the form (\ref{a1}). In
addition they have another pair of qubits in EPR state (\ref{a2}).
Alice (resp. Bob) has the first (resp. second) qubit of the EPR pair.
Therefore the initial state is given by
\begin{eqnarray*}
\vert \alpha\rangle \otimes \vert\beta\rangle  & =&
{1\over \sqrt 2}\bigl(a \ket 0\otimes(\ket {00} + \ket {11}) +
b\ket 1\otimes(\ket {00} +\ket {11})\bigr)\\  & =&
{1\over \sqrt 2}\bigl(a \ket {000} +a  \ket {011} + b \ket {100} +
b \ket {111}\bigr).
\end{eqnarray*}

Alice applies first the gate $C_{not}$ to her two qubits, then
the gate $H$ to the qubit whose state is to be teleported.
After these transformations the initial state becomes
\begin{eqnarray*}
\lefteqn{( H \otimes I \otimes I)(C_{not} \otimes
I)(\vert \alpha\rangle \otimes \vert \beta\rangle )} \\
&  = &  (H \otimes I \otimes
I){1\over \sqrt 2} \bigl(a\ket {000} + a\ket
{011} + b\ket {110} + b\ket {101}\bigr)\\
 &  = &  {1\over
2}\bigl(\ket {00} (a \ket {0} + b \ket {1}) +
         \ket {01} (a \ket {1} + b \ket {0})\\
&&+
         \ket {10} (a \ket {0} - b \ket {1}) +
         \ket {11} (a \ket {1} - b \ket {0})\bigr).
\end{eqnarray*}
Alice measures her two qubits. With equal probabilities she
will get one of the states: $\ket {00}$, $ \ket
{01}$, $ \ket {10}$ or $ \ket {11}$. Accordingly
Bob's qubit is projected to one of the states:
$ a \ket {0} + b \ket {1}$, $ a \ket {1} + b \ket {0}$, $
a \ket {0} - b \ket {1}$ or $ a \ket {1} - b \ket {0}$
respectively. Alice tells Bob the results of her measurement
by two classical bits information on classical channel. Bob
applies one of the gates: $I$, $X$, $Z$, or $Y$ with respect to
the classical bits he received.
$$
\begin{array}{ccc}
\mbox{bits received} &~~~~~~~  \mbox{state} ~~~~~~~&  \mbox{decoding} \\
 00 &  a \ket {0} + b \ket {1}&  I\\  01 &  a \ket
{1} + b \ket {0}&  X\\  10 &  a \ket {0} - b \ket {1}&
Z\\  11 &  a \ket {1} - b \ket {0}&  Y
\end{array}
$$
The state of Bob's qubit is then
transformed into $\vert\alpha\rangle$ exactly.

\section{Teleportation of general finite-dimensional quantum pure states}

The example in the preceding section is about the teleportation of a pure two
dimensional (qubit) state. We consider now the teleportation of a general
$N$-dimensional quantum state. We denote by $\{|i\rangle, i=0,...,N_{\theta}-1\}$ an
orthogonal normalized basis of an $N_\b$-dimensional Hilbert space
$H_{\theta}$, $\theta=1,2,3$. The spaces $H_1$ and $H_2$ are associated with Alice,
while $H_3$ is associated with Bob. Alice has particles in general
quantum states on the Hilbert space $H_1$ of the form
\be\label{psi0}
\vert \Psi_0\rangle=\left(\ba{c}\alpha_1\\ \vdots\\ \alpha_{N_1}
\ea \right)=\sum_{i=0}^{N_1-1}\alpha_i \vert i\rangle,~~~~\vert\Psi_0\rangle\in H_1,
\ee
where $\alpha_i\in \Cb$, $\sum_{i=0}^{N_1-1}\vert\alpha_i\vert^2=1$.

A general entangled state of two particles in the Hilbert spaces $H_2$
and $H_3$ is of the form
\be\label{psi1}
\vert \Psi_1\rangle =\sum_{i=0}^{N_2-1}\sum_{j=0}^{N_3-1}a_{ij} \vert ij \rangle,
~~~\sum_{i=0}^{N_2-1}\sum_{j=0}^{N_3-1}|a_{i
j}|^2=1
\ee
for some complex coefficients $a_{ij}\in\Cb$.
The degree of entanglement depends on the coefficients $a_{ij}$, $i=0,...,N_2-1$,
$j=0,...,N_3-1$. To send the state
$\vert\Psi_0\rangle$ to Bob's hand, it is necessary that $N_3\geq N_1$.
In the following we take $N_3=N_1$.

The initial state Alice and Bob have is then given by
\be\label{psi}
\vert \Psi_0\rangle \otimes \vert\Psi_1\rangle
=\sum_{i,k=0}^{N_1-1}\sum_{j=0}^{N_2-1}
\alpha_i a_{jk}\vert ijk \rangle
~~~~~~\in H_1\otimes H_2\otimes H_3\,.
\ee
Alice has the first and the second particles and Bob has the third one.
To transform the state of Bob's particle to be $\vert\Psi_0\rangle$,
similar to the qubit case, one has to do some
unitary transformation $U$ and measurements.
Let $U$  be the unitary transformation
acting on the tensor product of
two quantum states in the Hilbert spaces $H_1$ and $H_2$ such that
\be\label{u}
U (\vert ij \rangle)=\sum_{s=0}^{N_1-1}
\sum_{t=0}^{N_2-1}b_{ijst} \vert st \rangle,
\ee
with $\displaystyle\sum_{s=0}^{N_1-1}\sum_{t=0}^{N_2-1}
b_{ijst}b_{ijs^\prime t^\prime}^\ast=\delta_{ss^\prime}\delta_{tt^\prime}$,
$\forall i=0,1,...,N_1-1$, $j=0,1,...,N_2-1$.

\begin{Theorem}
If $b_{ijst}$ satisfies the following relation
\be\label{cond}
\sum_{i=1}^{N_1}\sum_{j=1}^{N_2}\alpha_i a_{jk} b_{ijst}=
\frac{1}{\sqrt{N_1N_2}}\alpha_{k-t+1}c_{s\,k-t+1\,t}
\ee
for some $c_{ijk}\in\Cb$ such that $c_{ijk}c_{ijk}^\ast=1$,
$U$ is the unitary transformation that fulfills the quantum teleportation.
\end{Theorem}

\noindent{\bf Proof}. From (\ref{u}) and (\ref{cond}) we have,
with $\vert \Psi_0\rangle$
(resp. $\vert\Psi_1\rangle$) as in (\ref{psi0}) (resp. (\ref{psi1})):
$$\ba{rcl}
(U\otimes 1)(\vert \Psi_0\rangle \otimes \vert\Psi_1\rangle)
\equiv\vert\psi\rangle&=&
\displaystyle\sum_{i,s,k=0}^{N_1-1}\sum_{j,t=0}^{N_2-1}
\alpha_i a_{jk}b_{ijst}\vert stk \rangle\\[4mm]
&=&\displaystyle\frac{1}{\sqrt{N_1N_2}}
\sum_{s,k=0}^{N_1-1}\sum_{t=0}^{N_2-1}
\alpha_{k-t+1} c_{s\,k-t+1\, t}\vert stk \rangle\\[4mm]
&=&\displaystyle\frac{1}{\sqrt{N_1N_2}}
\sum_{i,j=0}^{N_1-1}\sum_{k=0}^{N_2-1}
c_{ijk}\alpha_j \vert ik\,k+j-1 \rangle,
\ea
$$
where the indices $i,j,k$ in the basis vector $\vert ijk\rangle$
are understood to be taken
modulo by $N_1-1$, $N_2-1$ and $N_1-1$ respectively.

Now Alice measures her two qubits in the state $\vert\psi\rangle
\in H_1\otimes H_2$.
If $\vert ik \rangle$ is the state obtained after the
measurement, i.e.,
$$
\vert\psi\rangle\to \vert ik \rangle\otimes
\left(\sum_{j=0}^{N_1-1}c_{ijk}\alpha_j \vert k+j-1 \rangle\right),
$$
then in order to recover the original state $\vert\Psi_0\rangle$,
the unitary operator that Bob should use to act on his qubit is
\be\label{oik}
O_{ik}=P_k C_{ik},~~~~~~~i=0,1,...,N_1-1,~~~k=0,1,...,N_2-1,
\ee
where $P_k$ is the $(k-1)$-th power of the permutation operator,
$P_k=\Pi^{k-1}$,
$$
\Pi=\left(\ba{ccccc}
&&&&1\\
1&&&&\\
&1&&&\\
&&\ddots &&\\
&&&1&\ea
\right)
$$
(where the places without integer $1$ are zero)
and $C_{ik}=diag(c_{i1k}^\ast,c_{i2k}^\ast,...,c_{iN_1 k}^\ast)$.
After this transformation, one gets $\vert\psi\rangle\to \vert i k\rangle
\otimes\vert\Psi_0\rangle$ and the state $\vert\Psi_0\rangle$ given by
(\ref{psi0}) is
teleported from Alice to Bob. \hfill $\Box$

As a consequence of the above Theorem whenever an
entangled state in the sense of (\ref{psi1})
is given, i.e. the $a_{ij}$ are given, if there
are solutions of $b_{ijst}$ to equation (\ref{cond}), we have a unitary
transformation $U$ that fulfills the teleportation. The condition (\ref{cond})
can also be rewritten as
\be\label{cond1}
\sqrt{N_1 N_2}\sum_{i=0}^{N_2-1}a_{i\,t+j-1}b_{jist}=c_{sjt}\,.
\ee
The unitary transformation (\ref{u}) given by the
quantities $b_{jist}$ used in our teleportation protocol
depends on the initially given entangled state (\ref{psi1}) and the
dimensions of the Hilbert spaces $H_1$, $H_2$, $H_3$.

For general $N\equiv N_1=N_2=N_3$, if we take
\be\label{aij}
a_{ij}=\frac{\delta_{ij}}{\sqrt{N}},
\ee
the entangled state (\ref{psi1})
is given by
$$
\vert \Psi_1\rangle_{max}=\frac{1}{\sqrt{N}}\sum_{i=0}^{N-1}\vert ii \rangle.
$$
From equation (\ref{cond1}), we
obtain the unitary transformation (\ref{u}) used in the
teleportation protocol with
\be\label{bb}
b_{i\,t+i-1\, st}=\frac{c_{sit}}{\sqrt{N}},
\ee
with $c_{sit}$ as in (\ref{cond}),
the other coefficients $b$ in (\ref{u}) being zero.
It is easily checked that the transformation (\ref{u}) given by (\ref{bb})
is a unitary one. The teleportation is accomplished by applying the unitary
operation (\ref{oik}) according to the result of Bob's measurement.
For some particular values of the coefficients $c_{sit}$, this
result concide with the one in [11].

According to the Schmidt decomposition, see e.g. [18],
in this case the entangled state
(\ref{psi1}) on the Hilbert spaces $H_2$ and $H_3$ can be written as
$$
\sum_{i=0}^{N-1} \sqrt{\lambda_i} \vert ii \rangle
$$
in a suitable basis, where $\lambda_i\ge 0$, $\sum_{i=0}^{N-1}\lambda_i=1$.
That is, we can take $a_{ij}=\sqrt{\lambda_i}\delta_{ij}$. Substituting
this into equation (\ref{cond1}), we have
$$
\sqrt{\lambda_{t+j-1}}b_{j\,t+j-1\, st}=c_{sit}.
$$
According to the unitarity of the transformation $U$ and
the condition $c_{ijk}c_{ijk}^\ast=1$,
one gets $\lambda_i=1/N$, $i=0,1,...,N-1$, and the state
(\ref{psi1}) is maximally entangled, which shows that
with a less than maximally entangled state it is
impossible to give a unitary transformation that fulfills perfect
teleportations.

For $N=2$, taking
$c_{111}=c_{211}=c_{121}=c_{112}=c_{212}=c_{122}=
-c_{221}=-c_{222}=1$ (this choice satisfies the condition
$c_{ijk}c_{ijk}^\ast=1$), we have
$O_{11}=I$, $O_{12}=\sigma_x$, $O_{21}=\sigma_z$, $O_{22}=i \sigma_y$,
where $\sigma_{x,y,z}$ are Pauli matrices and $I$ is the $2\times 2$
identity matrix. The unitary transformation $U$ is then equal to the joint
actions of the controlled-not gate $C_{NOT}$ and the Walsh-Hadamard
transformation $H$, as defined, e.g., in [20,21].
This recovers the usual protocol
for teleporting two level quantum states \cite{Bennett93}, the most simple example
of teleportation given in section 2.

When $N=2^l$ for some $l\in\Nb$, a case discussed in [7],
$\vert\Psi_1\rangle$ can be rewritten as
$$
\vert\Psi_1\rangle=\prod_{i=1}^{l}\vert \beta\rangle_i
=\prod_{i=1}^{l}\frac{1}{\sqrt{2}}
(\vert 00\rangle+\vert 11\rangle)_i,
$$
$\vert \beta\rangle_i$ stands for the $i$-th EPR pair
with the first (resp. second) qubit attached to the
Hilbert space $H_2$ (resp. $H_3$).
Therefore instead of a fully entangled
state of two $N$-level qubits, we only need $l$ pairs of
entangled two-level qubits. This conforms with the
discussions in [7].

Generally, the dimension $N_2$ of $H_2$ can be greater than $N_1$.
As long as one prepares the entangled state of two qubits in
the Hilbert spaces  $H_3$ and the
sub Hilbert space ${\cal H}_2\subset H_2$, with dim(${\cal H}_2)=N_1$, the
above results are still valid.

We consider now some special cases of teleportations when some
components of the initial state are zero. Without  losing generality,
let $\alpha_i\neq 0$ for $i=1,...,n_1$, $n_1<N_1$,
and $\alpha_i=0$ for $i=n_1+1,...,N_1$
(We remark that for a given
$N_1$-dimensional vector it is always possible to make
some of its components
to be zero by changing the basis. But such a
basis transformation depends of course on the
components of the given vector, hence for an unknown quantum state
this kind of transformation has no practical use).

The initial state to be teleported under the above hypothesis
can be written as
$$
\vert \Psi_0\rangle
=\sum_{i=1}^{n_1}\alpha_i \vert i\rangle \,.
$$
We take the dimension of $H_2$ to be $N_2=n_1<N_1$.
The entangled state used to teleport $\vert\Psi_0\rangle$ can
be prepared in the following way:
\be\label{aij1}
a_{ij}=\left \{ \ba{l}
\displaystyle
\frac{1}{\sqrt{n_1}}\,\delta_{ij},~~~~~~~j=1,...,n_1\\[4mm]
0,~~~~~~~~~~~~~~~j=n_1+1,...,N_1
\ea\right.
\ee
for $i=1,2,...n_1$. From (\ref{cond1}) we get
the unitary transformation (\ref{u}) with
$$
b_{i\,t+i-1\, st}=\frac{c_{sit}}{\sqrt{N_1}}
$$
for $t, t+i-1~({\rm mod~n_1})=1,...,n_1$, $i,s=0,1,...,N_1-1$,
the other coefficients $b_{jist}$ in (\ref{cond1}) being zero.

An example is the teleportation of an EPR pair
$\vert\Psi_0\rangle=a\vert 01\rangle+b\vert
10\rangle$, $\vert a\vert^2 + \vert b\vert^2=1$,
as discussed in [4].
In this case we have $N_1=4$. $\vert\Psi_0\rangle$ can be written as
$\alpha \vert 3\rangle+\beta \vert 2\rangle
\equiv \alpha \vert 1^\prime\rangle+\beta \vert 2^\prime\rangle$.
The dimension of the auxiliary Hilbert space $H_2$ is only needed to be
$n_1=2$. The entangled state is given by $\vert\Psi_1\rangle=
\frac{1}{\sqrt{2}}(\vert 1 1^\prime\rangle+\vert 2 2^\prime\rangle)
=\frac{1}{\sqrt{2}}(\vert 1 3\rangle+\vert 2 2\rangle)
=\frac{1}{\sqrt{2}}(\vert 101\rangle+\vert 010\rangle)$.

Here as $n_1=N_1/2=2$, instead of (\ref{aij1}), we may alternatively
take $a_{ij}=\frac{1}{\sqrt{n_1}}\,\delta_{ij}$ for $j=n_1+1,...,N_1$ and
$a_{ij}=0$ for $j=1,...,n_1$. Then the entangled state becomes
$\vert\Psi_1\rangle=\frac{1}{\sqrt{2}}(\vert 11\rangle+\vert 24\rangle)
=\frac{1}{\sqrt{2}}(\vert 000\rangle+\vert 111\rangle)$, which is called
a GHZ triplet consisting of three two-level qubits and can be realized
experimently [22,23]. The unitary transformation is given by
$b_{i\,t+i-1\, st}=c_{sit}/\sqrt{N_1}$
for $t=1,...,n_1$, $t+i-1~({\rm mod~n_1})=n_1+1,...,N_1$, $i,s=0,1,...,N_1-1$,
and the other coefficients $b_{jist}$ in (\ref{cond1}) being zero. For a
suitable choice of the sign of $c_{sit}$,
this recovers the result in [4].

\section{Standard teleportation of mixed states}

In realistic situations, due to the
interactions with the environment, instead of a
pure state, Alice may have a mixed state $\r$. And
instead of the pure maximally entangled states, Alice and Bob
usually share a mixed entangled state $\chi$. Before considering the
the teleportation of a mixed state by a mixed entangled state
as an entangled resource, we first reformulate the teleportation
protocol for pure states.

Any linear operator $A:~ H_{\a}\longrightarrow H_{\b}$
can be represented by an $N_{\b}\times N_{\a}$-matrix as follow:
\bea
A(|\a\rangle)=\sum_{i=0}^{N_{\b}-1}A_{\a i}|i\rangle,~~~~~~\vert\a\rangle\in
H_{\a}.\no
\eea
Take $N_1=N_3=n$, $N_2=m$. To transform the state
$|\P_0\rangle$ in (\ref{psi0}) to Bob by using the entangled state (\ref{psi1})
and the unitary transformation $U$ in (\ref{u}), we
introduce a number $mn$ of $m\times n$-matrices with matrix
elements $(B_{st})_{ij}=b_{ij,st}$. Let $A$ be an $n\times m$ matrix
with matrix element $(A)_{i j}=a_{i j}$. We have

\begin{Theorem} If $\lt\{B_{st}\rt\}$ and $A$ satisfy the following
relation
\bea
\lt\{\begin{array}{l}
tr \lt(B_{st}B^+_{s't'}\rt)=\d_{tt'}\d_{ss'}\,,\\
B_{st}^+A^+A~B_{st}=|\l_{st}|^2~I_{n\times n}\,,\\
AA^+=\frac{1}{n}I_{n\times n}\,,
\end{array}\rt.\label{Mai}
\eea
for some nonzero complex number $\l_{st}$ such that
$\sum_{s,t}|\l_{st}|^2=1$, then $U$ is a unitary transformation
that fulfills perfect quantum teleportation.
\end{Theorem}

\noindent{\bf Proof}. The first condition in (\ref{Mai}) is equal
to the unitary condition for the transformation (\ref{u}). From
(\ref{psi}) and (\ref{u}), after Alice measures her two
particles with outcoming state $|st\rangle$, Bob's particle will become
\bea
|\P_0\rangle\longrightarrow T_{st}|\P_0\rangle=AB_{st}|\P_0\rangle.\no
\eea
If the condition (\ref{Mai}) is satisfied, then we have
$T^{+}_{st}T_{st}=T_{st}T^{+}_{st}=|\l_{st}|^2I_{n\times n}$, and
perfect quantum teleportation is fulfilled. \hfill $\Box$

For the $m=n$ case, introducing $U_{st}=\sqrt{n}B_{st}$, condition
(\ref{Mai}) is equivalent to
\bea
\lt\{\begin{array}{l}
tr \lt(U_{st}U^+_{s't'}\rt)=n \d_{tt'}\d_{ss'},\\
U_{st}~U_{st}^+=I_{n\times n},
\end{array}\rt.\label{Wer}
\eea
with
\bea AA^+=A^+A=\frac{1}{n}I_{n\times n} ~{\rm and}
~~|\l_{st}|^2=\frac{1}{n^2}.
\eea
Hence, only the maximally
entangled state (\ref{psi1}) (satisfying
$AA^+=A^+A=\frac{1}{n}I_{n\times n}$) can fulfill the perfect quantum
teleportation. For the maximally entangled state shared by Alice
and Bob, in order to teleport an unknown state perfectly, one
should find a number $n^2$ of $n\times n$-matrices which satisfy (\ref{Wer})
(these constitute, according to Werner's definition \cite{werner}, a basis for
the unitary operators). The classification of the general solution
of (\ref{Wer}) is an open problem \cite{werner}. Fortunately, we
can construct a special solution up to a global unitary
transformation in the sense of [8] as follows.

Let the $n\times n$ matrices $h,~g$ be such that
$h|j\rangle=|(j+1)~mod~n\rangle,~~g|j\rangle=\o^j|j\rangle$, with
$\o=exp\{\frac{-2i\pi}{n}\}$. Define $n^2$ linear-independent
$n\times n$-matrices
\be\label{ust}
U_{st}=h^{t}g^s
\ee
which satisfy
\bea
U_{st}U_{s't'}=\o^{st'-ts'}
U_{s't'}U_{st},~~tr(U_{st})=n\d_{s0}\d_{t0}. \label{Mat}
\eea
One can check that such $U_{st}$ in (\ref{Mat}) satisfy equation
(\ref{Wer}) and form a complete basis of $n\times n$-matrices.
They  can be used to complete the perfect teleportation through a
maximally entangled state \cite{Bennett93,Bra00}.

In the following part of this paper, we focus on the case $m=n$.
Set
\be\label{Phimax}
|\Q\rangle=\frac{1}{\sqrt{n}}\sum_{i=0}^{n-1}|ii\rangle,
\ee
and let the $\{U_{st}\}$ be as in (\ref{Mat}). The standard  teleportation
protocol $T_0$ \cite{Bennett93} can be written as the following linear
transformation \be\ba{rcl} {\cal T}_{T_0}^{\lt\{|\Q\rangle\rt\}}(|\p\rangle)
&=&\sum_{s,t}(\frac{1}{\sqrt{n}}B_{st})^+
\lt( P_{st}\lt(U\otimes 1\lt(|\p\rangle\otimes |\Q\rangle\rt)\rt)\rt)\\[3mm]
&=&\sum_{s,t}\frac{1}{\sqrt{n}}B_{st}^+\frac{1}{\sqrt{n}}B_{st}|\phi\rangle=|\phi\rangle,
\ea\ee where $P_{st}$ is the projection $(|st\rangle\langle st|)\otimes 1$.
Namely, the noiseless quantum channel is \bea
\L_{T_0}^{(|\Q\rangle\langle\Q|)}(\r)=\r. \eea

Although in principle one can create pure and maximally entangled
states for teleportation, in a realistic situation any pure state
will be evolved into a mixed state due to its interaction with the
environment (decoherence). These unwanted interactions show up as
noise in quantum information processing systems. Teleportation
using a mixed state as an entangled  resource is, in general,
equivalent to having a noisy quantum channel.

We shall derive
an explicit expression for the quantum channel associated with
the standard teleportation protocol in terms of an arbitrary mixed
state resource. From the complete basis (\ref{Mat}), one can
introduce a complete orthogonal normalized basis $\{|\Q_{st}\rangle\}$
of the Hilbert spaces associated with Alice and Bob,
\bea
|\Q_{st}\rangle=(1\otimes
U_{st})|\Q\rangle=\frac{1}{\sqrt{n}}\sum_{i,j}(U_{st})_{ij}|ij\rangle ,~~~{\rm
and }~~|\Q_{00}\rangle=|\Q\rangle.\label{Max-sta}
\eea
As locally unitary transformations do not change the degree of the
entanglement, each $|\Q_{st}\rangle$ is a maximally entangled state.

\begin{Theorem}
The standard teleportation protocol $T_0$, when used with an arbitrary mixed
state with
density matrix $\x$, as a resource, acts as a general trace-preserving quantum
channel,
\bea
\L_{T_0}^{(\x)}(\r)=\sum_{s,t=0}^{n-1}\langle\Q_{st}|\x|\Q_{st}\rangle U_{st}~\r
~ U_{st}^+\,.\label{Res}
\eea
\end{Theorem}

\noindent{\bf Proof}. If Alice and Bob share with a general pure
state (\ref{psi1}) as their quantum resource, after the measurement
based on the Bell-basis $|\Q_{st}\rangle$, Bob's
particle becomes an (unnormalized) state
\bea
|\phi\rangle\longrightarrow
\lt( P_{st}\lt(U\otimes 1\lt(|\phi\rangle\otimes |\P\rangle\rt)\rt)\rt)
=AB_{st}|\phi\rangle.\no
\eea
In terms of the density matrix, we have
\bea
\r\longrightarrow AB_{st}~\r~ B_{st}^+A^+\,.\no
\eea
After some
local unitary operations by Bob, according to the results
of the measurement made by Alice, the final state can be given as
\bea
\L^{(|\P\rangle\langle\P|)}(\r)=\sum_{t,\b}U_{t\b}^{+}AB_{t\b}~\r~
B_{t\b}^+A^+U_{t\b}.\no \eea

Suppose that $\lt\{p_{\a}, |\P_{\a}\rangle\rt\}$ is one of the ensemble
constituting the arbitrary mixed state $\x$, i.e.
\be\label{mixed}
\x=\sum_{\a}p_{\a}|\P_{\a}\rangle\langle\P_{\a}|,~~0\leq p_{\a}\leq 1
\ee
and
$$
\sum_{\a}p_{\a}=1,~~|\P_{\a}\rangle=\sum_{i,j}a^{(\a)}_{ij}|ij\rangle.
$$
From the standard teleportation protocol $T_0$, with the
resource state $\x$, the final Bob's state becomes
\be\label{24}
\L^{(\x)}_{T_0}(\r)=\sum_{t,\b}\sum_{\a}p_{\a}U^{+}_{t\b}A^{(\a)}B_{t\b}~\r~
B^{+}_{t\b}(A^{(\a)})^+U_{t\b}.
\ee

Since each matrix
$A^{(\a)}$ can be decomposed in the basis $\lt\{U_{st}\rt\}$,
\be\label{add}
(A^{(\a)})_{ij}=\sum_{s,t}a^{(\a)}_{st}(U_{st})_{ij},
\ee
using $B_{t\b}=\frac{1}{\sqrt{n}}U_{t\b}$, (\ref{24}) becomes
\bea
\L^{(\x)}_{T_0}(\r)&=&\frac{1}{n}\sum_{s,t}\sum_{s',t'}\lt(
\sum_{\a}p_{\a}a^{(\a)}_{st}a^{(\a)*}_{s't'}\rt)\sum_{\g,\b}
U_{\g\b}^+U_{st}U_{\g\b}~\r~ U^+_{\g\b}U_{s't'}^+U_{\g\b}\no\\
&=&\frac{1}{n}\sum_{s,t}\sum_{s',t'}\lt(
\sum_{\a}p_{\a}a^{(\a)}_{st}a^{(\a)*}_{s't'}\rt)U_{st}~\r~
U_{s't'}^+\sum_{\g\b}\o^{s\b-t\g-s'\b+t'\g}\no\\
&=&\frac{1}{n}\sum_{s,t}\sum_{s',t'}\lt(
\sum_{\a}p_{\a}a^{(\a)}_{st}a^{(\a)*}_{s't'}\rt)U_{st}~\r~
U_{s't'}^+~n^2\d_{ss'}\d_{tt'},\label{xyz}
\eea
where we have used
the equations (\ref{Mat}) and the identity
$\sum_{k=1}^{n}\o^{mk}=n\d_{m0}$. Using the definition of the
generalized Bell-states $\lt\{|\Q_{st}\rangle\rt\}$ given in
(\ref{Max-sta}), after a lengthy calculation, we arrive at
\bea
n\sum_{\a}p_{\a}a^{(\a)}_{st}a^{(\a)*}_{s't'}=\langle\Q_{st}|\x|\Q_{s't'}\rangle.\no
\eea
Substituting the above results into (\ref{xyz}), one obtains
(\ref{Res}). Using (\ref{Wer}), the trace-preserving property of
the quantum channel can be derived from the following useful
identity
\bea
\sum_{s,t}U_{st}~\r~U_{st}^+=n~tr(\r)~I_{n\times
n}.\label{Com}
\eea
\hfill$\Box$

The most important consequence of our result is that in order to
fulfill the perfect noise-free teleportation, the entangled
resource should not only be a maximally entangled state but also a
pure state \cite{werner}, otherwise the out coming state is always mixed.
The success of teleportation can be quantified by the transmission
fidelity between outstate and instate defined by
\bea
f(\x)=\overline{\langle\phi_{in}|
\L_{T_0}^{(\x)}(|\phi_{in}\rangle\langle\phi_{in}|)|\phi_{in}\rangle},\label{Fed}
\eea
averaged over all pure input states $\phi_{in}$.

In order to calculate the transmission fidelity (\ref{Fed}), we
need an elementally irreducible representation {\bf G} of the unitary
group ${\bf U(n)}$. Let $U(g)$ be the unitary
matrix representation of any element of {\bf G}. Recalling Schur's
Lemma, one has the identity \bea &&\int_{{\bf G}}
dg~(U^+(g)\otimes U^+(g))~\s~(U(g)\otimes U(g))=\a_1I\otimes
I+\a_2 P,
\label{Shu}\\
&&\a_1=\frac{n^2tr(\s)-ntr(\s P)}{n^2(n^2-1)},~~
\a_2=\frac{n^2tr(\s P)-ntr(\s )}{n^2(n^2-1)},\no \eea for any
operator $\s$ acting on the tensor space, where $P$ is the flip
operator such that $P|ij\rangle=|ji\rangle$. The invariant (Haar) measure $dg$
on ${\bf G}$ is normalized such that $\int_{{\bf G}}dg=1$.

\begin{Theorem}
The transmission fidelity of the standard teleportation protocol
with arbitrary mixed state $\x$ as a resource is given by
\bea
f(\x)=\frac{n}{n+1}F(\x)+\frac{1}{n+1},\label{Fe}
\eea
where
$F(\x)=\langle\Q|\x|\Q\rangle$ is the singlet fraction of the resource $\x$.
\end{Theorem}
\vspace{0.4truecm}

\noindent{\bf Proof}. From Theorem 3, we have
\bea
f(\x)&=&\sum_{s,t}\langle\Q_{st}|\x|\Q_{st}\rangle\overline{\langle\phi_{in}|
U_{st}|\phi_{in}\rangle\langle\phi_{in}|U^+_{st}|\phi_{in}\rangle}\no\\
&=&\sum_{s,t}\langle\Q_{st}|\x|\Q_{st}\rangle\overline{\langle\phi_{in}|\langle\phi_{in}|~
(U_{st}\otimes U^+_{st})~|\phi_{in}\rangle|\phi_{in}\rangle}\no\\
&=&\sum_{s,t}\langle\Q_{st}|\x|\Q_{st}\rangle\\
&&\int_{{\bf G}}dg~\langle00|
(U^+(g)\otimes U^+(g))~ (U_{st}\otimes U^+_{st})~(U(g)\otimes U(g))~|00\rangle\no\\
&=&\frac{1}{n(n+1)}\sum_{s,t}\langle\Q_{st}|\x|\Q_{st}\rangle\lt\{tr(U_{st})~tr(U^+_{st})
+tr(U_{st}U^+_{st})\rt\}\no\\
&=&\langle\Q_{00}|\x|\Q_{00}\rangle\frac{n}{n+1}+\frac{1}{n+1},\no
\eea
where we have used the identity (\ref{Shu}) and $tr_{12}\lt((A\otimes
B)P\rt)=tr(AB)$ in deriving the fourth equation, and have used the identity
(\ref{Mat}) in the last equation. \hfill $\Box$

Formula (\ref{Fe}) implies that the transmission fidelity of the
standard teleportation protocol depends on the maximally entangled
fraction only. Hence, in order to improve the transmission
fidelity of the standard teleportation protocol with a given
entangled resource, one must distill the singlet fraction of the
resource in terms of a distillation protocol \cite{BBPSSW}.

\section{Optimal teleportation of mixed quantum states}

From above we see that the main operations in the quantum
teleportation protocol are: 1) a unitary transformation and a
measurement by Alice (Bell measurement); 2) according to the
results of the measurement, a unitary transformation by Bob. In
the last section the unitary transformation used by Bob is the
same as the one in standard teleportation (i.e., when the
entangled state used in teleportation is maximally entangled). A
question one would ask is wether the fidelity given by (\ref{Fe})
is the best one or not. In this section we consider the following
problem: based on the Bell measurement, what kind of unitary
transformation should be used by Bob to get an optimal fidelity.

$\lt\{U_{st}\rt\}$ in (\ref{ust}) form a complete basis of
$n\times n$-matrices, namely, for any $n\times n$ matrix $W$, $W$
can be expressed as
\bea
W=\frac{1}{n}\sum_{s,t}tr
(U_{st}^+W)U_{st}.\label{Dec}
\eea
In general, all
the maximally entangled pure states are equivalent to $|\Q\rangle$
in (\ref{Phimax}):
$|\Psi_{max}\rangle=1\otimes U|\Q\rangle$, where $U$ is a unitary
transformation. One can define the fully entangled fraction
\cite{Hor01} of a state $\x$ by
\be\label{MF1}
\F(\x)=max\lt\{\langle\Q|(1\otimes
U^+)~\x~(1\otimes U)|\Q\rangle\rt\}
\ee
for all $UU^+=U^+U=I_{n\times n}$. Since the group of
unitary transformations in n-dimensions is compact, there exists
an unitary matrix $W_{\x}$ such that
\bea
\F(\x)=\langle\Q|(1\otimes
W_{\x}^+)~\x~(1\otimes W_{\x})|\Q\rangle.\label{MF2}
\eea

Suppose now again Alice and Bob previously share a pair of particles in
an arbitrary mixed entangled state $\x$. To transform  an unknown
state  to Bob, Alice first performs a joint Bell measurement based
on the generalized Bell-states (\ref{Max-sta}) on her parties.
According to the measurement results of Alice, Bob chooses
particular unitary transformations $\lt\{T_{st}\rt\}$ to act on
his particle.

\begin{Theorem}
The teleportation protocol defined by $\lt\{T_{st}\rt\}$, when
used with an arbitrary mixed state with density matrix $\x$ as a
resource,  acts as a general trace-preserving quantum channel
\be\ba{rcl}
\L^{(\x)}_{\{T\}}(\r)&=&\displaystyle\frac{1}{n^2}\sum_{s,t}\sum_{s',t'}
\langle\Q_{st}|\x|\Q_{s't'}\rangle\\[5mm]
&&\lt\{\sum_{\g\b}
T_{\g\b}^+U_{st}U_{\g\b}~\r~
U^+_{\g\b}U_{s't'}^+T_{\g\b}\rt\}.\label{Out}
\ea
\ee
\end{Theorem}

\noindent{\bf Proof}. The proof can be given in two steps:

\noindent{\it Step 1. Pure entangled state as a resource}.
To transform the state $|\P_0\rangle$ to Bob, Alice
performs a joint Bell measurement based on the generalized
Bell-states Eq.(\ref{Max-sta}) on her party. After her measurement
with outcoming in the state $|\Q_{st}\rangle$, Bob's particle gets into
an (unnormalized) state
\bea
|\p\rangle\longrightarrow
\frac{1}{\sqrt{n}}AU_{st}|\p\rangle.\no
\eea
Once Bob learns from Alice
that she has obtained the result $st$, he performs on his
previously entangled particle (particle 3) a unitary
transformation $T_{st}$. Then the final state becomes
$\frac{1}{\sqrt{n}}T^+_{st}AU_{st}|\p\rangle$. In terms of the density
matrix, the teleportation based on the unitary matrices $\left\{
T_{st}\right\}$ , with quantum resource being a pure state $|\P_1 \rangle$,
is a quantum channel with the output
\bea
\L^{(|\P_1\rangle\langle\P_1|)}_{\{T\}}(\r)=\frac{1}{n}\sum_{st}T^+_{st}AU_{st}~\r~
U^+_{st}AT_{st}.\no
\eea

\noindent{\it Step 2. An arbitrary mixed entangled state as a
resource.}
Applying the teleportation protocol $T$ with a mixed state
$\x$, Bob's state becomes
\bea
\L^{(\x)}_{\{T\}}(\r)=\frac{1}{n}\sum_{s,t}\sum_{\a}p_{\a}T^{+}_{st}A^{(\a)}
U_{st}~\r~U^{+}_{st}(A^{(\a)})^+T_{st}.\label{Tel}
\eea
As $A^{(\a)}$ can be decomposed in the basis $\lt\{U_{st}\rt\}$ by
formula (\ref{add}), (\ref{Tel}) becomes
\bea
\L^{(\x)}_{\{T\}}~(\r)&=&\frac{1}{n}\sum_{s,t}\sum_{s',t'}\lt(
\sum_{\a}p_{\a}a^{(\a)}_{st}a^{(\a)*}_{s',t'}\rt)\sum_{\g,\b}
T_{\g\b}^+U_{st}U_{\g\b}~\r~ U^+_{\g\b}U_{s't'}^+T_{\g\b}.\no
\eea
Using the definition of generalized Bell-states
$\lt\{|\Q_{st}\rangle\rt\}$ in (\ref{Max-sta}), after a lengthy
calculation, we arrive at
\bea
n\sum_{\a}p_{\a}A^{(\a)}_{st}A^{(\a)*}_{s',t'}=\langle\Q_{st}|\x|\Q_{s't'}\rangle.
\no
\eea

Substituting the above results into (\ref{Tel}), we
obtain (\ref{Out}). Using (\ref{Wer}) and the identity
\bea
\sum_{s,t}U^+_{st}~A~U_{st}=n\, tr(A)\, I_{n\times n},~~{\rm for
~any~} n\times n ~{\rm matrix}~ A,\no
\eea
the trace-preserving
property of  the quantum channel can be proved,
\bea
tr\lt(\L^{(\x)}_{\{T\}}(\r)\rt)
&=&\frac{1}{n^2}\sum_{s,t}\sum_{s',t'}
\langle\Q_{st}|\x|\Q_{s't'}\rangle\lt\{\sum_{\g\b}~tr\lt(
U^+_{\g\b}U_{s't'}^+U_{st}U_{\g\b}\r \rt)\rt\}\no\\
&=&\frac{1}{n}\sum_{s,t}\sum_{s',t'}
\langle\Q_{st}|\x|\Q_{s't'}\rangle tr\lt(U_{st}U_{s't'}^+\rt)\times tr(\r)\no\\
&=&\sum_{s,t}\langle\Q_{st}|\x|\Q_{st}\rangle=tr(\x)=1.\no
\eea
\hfill $\Box$

\begin{Theorem}
The transmission fidelity of the teleportation protocol defined by
$\lt\{T_{st}\rt\}$ with arbitrary mixed state $\x$ as a resource
is given by
\be\ba{rcl}
f(\x)&=&\displaystyle\frac{1}{n(n+1)}\sum_{\g\b}\langle\Q|\lt(1\otimes
(T_{\g\b}U^{+}_{\g\b})^+\rt)~\x~ \lt(1\otimes
T_{\g\b}U^+_{\g\b}\rt)|\Q\rangle\\[5mm]
&&+\displaystyle\frac{1}{n+1}.
\ea
\ee
\end{Theorem}

\noindent{\bf Proof}.
From (\ref{Fed}), Theorem 5 and formula (\ref{Shu}), one has
$$
\ba{rcl}
f(\x)
&=&\displaystyle\frac{1}{n^2}\sum_{s,t}\sum_{s',t'}\langle\Q_{st}|\x|\Q_{s't'}\rangle\\[4mm]
&&\sum_{\g\b}\overline{\langle\phi_{in}|
T^+_{\g\b}U_{st}U_{\g\b}|\phi_{in}\rangle\langle\phi_{in}|U^+_{\g\b}U^+_{s't'}T_{\g\b}
|\phi_{in}\rangle}\no\\[4mm]
&=&\displaystyle\frac{1}{n^2}\sum_{s,t}\sum_{s',t'}\langle\Q_{st}|\x|\Q_{s't'}\rangle\\[4mm]
&&\sum_{\g\b}\overline{\langle\phi_{in}|\otimes\langle\phi_{in}|\lt(
T^+_{\g\b}U_{st}U_{\g\b}\otimes U^+_{\g\b}U^+_{s't'}T_{\g\b}\rt)
|\phi_{in}\rangle\otimes|\phi_{in}\rangle}\no\\[4mm]
&=&\displaystyle\frac{1}{n^2}\sum_{s,t}\sum_{s',t'}\langle\Q_{st}|\x|\Q_{s't'}\rangle
\sum_{\g\b}\langle00| \int_{{\bf G}} dg~(U(g)^+\otimes U(g)^+)\no\\[4mm]
&&\lt(
T^+_{\g\b}U_{st}U_{\g\b}\otimes U^+_{\g\b}U^+_{s't'}T_{\g\b}\rt)(U(g)\otimes U(g))
|00\rangle\no\\[4mm]
&=&\displaystyle\frac{1}{n^3(n+1)}\sum_{s,t}\sum_{s',t'}\langle\Q_{st}|\x|\Q_{s't'}\rangle\\[4mm]
&&\sum_{\g\b}\lt\{
tr\lt(T^+_{\g\b}U_{st}U_{\g\b}\rt)tr\lt(
U^+_{\g\b}U^+_{s't'}T_{\g\b}\rt)\rt.\\[4mm]
&&\lt.+tr\lt(T^+_{\g\b}U_{st}U_{\g\b}
U^+_{\g\b}U^+_{s't'}T_{\g\b}\rt)\rt\}\no\\[4mm]
&=&\displaystyle\frac{1}{n(n+1)}\sum_{\g\b}\langle\Q|\lt(1\otimes
(T_{\g\b}U^{+}_{\g\b})^+\rt)\x \lt(1\otimes
T_{\g\b}U^+_{\g\b}\rt)|\Q\rangle+\frac{1}{n+1},
\ea
$$
where the identity $tr_{12}\lt((A\otimes B)P\rt)=tr(AB)$, (\ref{Wer}) and
(\ref{Dec}) have been used. \hfill $\Box$

Obviously when the term $\langle\Q|\lt(1\otimes
(T_{\g\b}U^{+}_{\g\b})^+\rt)\x \lt(1\otimes
T_{\g\b}U^+_{\g\b}\rt)|\Q\rangle$ is maximized, i.e.,
$T_{\g\b}U^{+}_{\g\b}=W_\x$, one gets the maximal fidelity.
Recalling the definition of the fully entangled fraction
(\ref{MF1}) and (\ref{MF2}), we arrive at our main result:

\begin{Theorem}
The optimal teleportation based on the Bell measurements, when
used with an arbitrary mixed state with density matrix $\x$ as a
resource, acts as a general trace-preserving quantum channel
\be
\ba{rcl}
\L^{(\x)}_{O}~(\r)&=&\frac{1}{n^2}\sum_{s,t}\sum_{s',t'}
\langle\Q_{st}|\x|\Q_{s't'}\rangle\\[5mm]
&&\lt\{\sum_{\g\b} U_{\g\b}^+W_{\x}^+U_{st}U_{\g\b}~\r~
U^+_{\g\b}U_{s't'}^+W_{\x}U_{\g\b}\rt\}.
\ea
\ee
The corresponding transmission fidelity is given by
\bea
f_{max}(\x)=\frac{n\F(\x)}{n+1}+\frac{1}{n+1},
\eea
where $\F(\x)$
is the fully entangled fraction (\ref{MF1}) and $W_{\x}$ is the
unitary matrix which fulfills such a fully entangled fraction
in (\ref{MF2}).
\end{Theorem}

We have studied the general properties of
teleportation for finite dimensional discrete pure quantum states.
The protocol we presented is for generally given entangled states with
$N_3=N_1$. If $N_3>N_1$, one can always take a subspace ${\cal
H}_3\subset H_3$ such that
dim$({\cal H}_3)=N_1$ and prepare the entangled state in the Hilbert spaces
$H_2$ and ${\cal H}_3$. Accordingly the initial state $\vert\Psi_0\rangle$
will be sent to the subspace ${\cal H}_3$.

For teleportation of quantum mixed states, we obtained
an explicit expression of the output
state of the optimal teleportation, with arbitrary mixed entangled
state as resource,  in terms of some noisy quantum channel. This
allows us to calculate the transmission fidelity of the quantum
channel. It is shown that the transmission fidelity depends only
on  the fully entangled fraction of the quantum resource
shared by the sender and the receiver, whereas that of a standard
teleportation depends on the singlet fraction.
Therefore the fidelity in our optimal
teleportation protocol is in general greater than the one in
the standard teleportation protocol \cite{Bennett93,Bow01,Alb02,Alb03}.

\vspace{1.0truecm}

\noindent {\bf Acknowledgments}
\noindent {\bf Acknowledgments}
W-L Yang would like to thank Prof. G. von Gehlen for his continous
encouragement during his visiting in Physikalisches Institut der
Universit\"at Bonn, where some part of this work was done. He also
would like to thank Prof. R.Sasaki and the Yukawa Institute for
Theoretical Physics, Kyoto University for their warm hospitality.
He is supported by the Japan Society for the Promotion of Science.

\end{document}